\documentclass[conference]{IEEEtran}
\IEEEoverridecommandlockouts
\usepackage{cite}
\usepackage{amsmath,amssymb,amsfonts}
\usepackage{algorithmic}
\usepackage{graphicx}
\usepackage{textcomp}
\usepackage{xcolor}
\def\BibTeX{{\rm B\kern-.05em{\sc i\kern-.025em b}\kern-.08em
    T\kern-.1667em\lower.7ex\hbox{E}\kern-.125emX}}
\begin{document}

\title{Population preferences through  Wikipedia edits\\}

\author{\IEEEauthorblockN{Y\'{e}rali Gandica}
\IEEEauthorblockA{ Center for Operations Research and Econometrics (CORE),\\
Institute of Information and Communication Technologies, Electronics and Applied Mathematics (ICTEAM)\\
Universit{\'e} catholique de Louvain, Louvain-la-Neuve, \\
and Center for Research in Finance and Management (CeReFiM), Universit\' e de Namur, Namur, \\
Belgium.\\
ygandica@gmail.com}
}

\maketitle

\begin{abstract}
In this work, we are interested in the inner-cultural background shaping broad people's preferences. Our interest is also to track this human footprint, as it has the tendency to
disappear due to the nowadays globalization. Given that language is a social construction, it is part of the historical reservoir, shaping the cultural (and hence collective) identity, then 
helping the community to archive accumulated knowledge about its culture and identity. We assume that the collective interest of a language-speaking community to document their events,
people and any feature important for them, by the online encyclopedia Wikipedia, can act as a footprint of the whole group's collective identity. The analysis of the language's preferences 
into categories among several languages, could have also applications into the field of Multilingual Natural Language Processing (MNLP). We, then, report results about the number of edits, editors, and pages into
categories, displayed by the several languages. Results are shown by several angles, and some extra measures complement the analysis.
\end{abstract}

\section{Introduction}
Wikipedia, the multilingual encyclopedia project, which is  supported by the Wikimedia Foundation, and based on a model of openly editable content, is available in 291 
languages. Even though there is no central authority who dictates which topics should be covered, one of the challenges for 
Wikipedia, in the last decade, has been to balance the coverage of content across its different languages \cite{West2016,omnipedia}. For this purpose, a recommendation system is 
applied by the Wikimedia Foundation in order to encourage Wikipedians to fill that gap \cite{Manske}.  
\\
\\
However, contributing to Wikipedia means more than writing encyclopedic contents. Indeed, it allows communities to store cultural memories of events, to show the reality by their
own lens and to document their prominent people and places \cite{samolienko}. In this sense, even though we understand the inconvenience of the imbalance between the information 
among several languages, we hypothesize that this collectively genuine gap has some important implications. It represents legitimate preferences among individuals sharing the 
same language, which is a footprint of the groups’ collective identity.
\\
\\
Our goal, in this communication, is to analyze the broad preferences of the population editing Wikipedia (WP), depicted by categories over several worldwide languages. In addition, 
we are also investigating cultural language-based footprints, since they have a tendency to disappear due to nowadays globalization. Our analysis is, hence, limited to the first 
10 years of the edits in each language, when no intervention to cover the gap between languages had yet been done. Our study covers twelve Wikipedias: the ones written in 
English (EN-WP), Spanish (ES-WP), French (FR-WP), Portuguese (PT-WP), Italian (IT-WP), Hungarian (HU-WP), German (DE-WP), Russian (RU-WP), Arabic (AR-WP), Japanese (JA-WP), 
Chinese (ZH-WP) and Vietnamese (VI-WP). Our selection has been done based on the interplay between a worldwide view and the WP sizes. Some limitations are present in our study, 
as the fact that to some extent, some WP languages have more global than local character, as for example the English one (EN-WP), which is worldwide edited. This language is only used 
for comparative purposes. 
\section{Methodology}

In order to have a broad view regarding preferences among individuals sharing the same language, we have chosen the categories already classified as in 
the main branch of the tree structure, defined by the Wikimedia Foundation itself, and found in \cite{tree}.  For the sake of avoiding a strong overlap between categories,
after taking a look at the pages within each category, our only intervention has been the decision of leaving out of the study the categories: Culture, 
Humanities‎, Law, Life, Matter, People, Reference Works, Science and Technology, Society, Universe, and World. Given that the pages within such categories appeared in several of them. 
Then, over the original 22 categories, our 
study is restricted to the 13 following ones: Arts, Sports, Right, Events, Philosophy, Geography, History, Games, Mathematics, Nature, Politics, Religion, 
and Health. We, then, have used the petscan API \cite{API} in order to download the names of all the pages within each category. This procedure has been done
for the 12 languages under study. 

In order to ensure the same category between languages, we started from the category in English (EN) and link
to the WP page for the same category in the desired language. Once the name of the category in the new language was obtained, then, all the pages within that category were collected, by 
fixing the category and the new language in the petscan API \cite{API}, without any link to the EN-WP. In this sense, different languages have different number of pages in each category.

All the activity of edits on the openly available Wikipedia was downloaded, also available in \cite{konect}. The starting point has been to set the starting date for each language, which ranges from
$11/10/2001$ to $28/03/2010$ for all of them. Because some information was not related to edits, but to other processes of Wikipedia's routines, only the pages fitting with the pages in one of 
the studied categories were taken into account. We have removed the activity of the bots, by removing all the edits done by users whose name contains the word bot in any combination 
of both uppercase and lowercase. In Table \ref{tab:twitternets} we show the number of pages studied and the number of edits for each language. This will constitute
the data-set of our study. 

%\section*{Tables}
\begin{table}[h!]
\centering
\begin{tabular}{|c|c|r|r|}
\hline
 WP & \# pages & \# edits \\
 \hline
 $EN$ & 1999263 & 222417366 \\
 $ES$ & 1144177 & 47728243  \\
 $FR$ & 2936383 & 58325545 \\
 $PT$ & 894521 & 19937771 \\
 $IT$ & 1084333 & 22200807  \\
 $HU$ & 248808 & 5758998  \\
 $DE$ & 1111265 & 39689676  \\
 $RU$ & 1134752 & 14199590 \\
 $AR$ & 624118 & 7674946 \\
 $JA$ & 690795 & 24584471 \\
 $ZH$ & 495855 & 3657770 \\
 $VI$ & 238859 & 12618296 \\
 \hline
\end{tabular}
\vspace{0.5cm}
\caption{Characteristics of the data-set.}
\label{tab:twitternets}
\end{table}

\section{Results}
\subsection{By languages}
\begin{figure}[tbp]
\scalebox{0.5}{\includegraphics{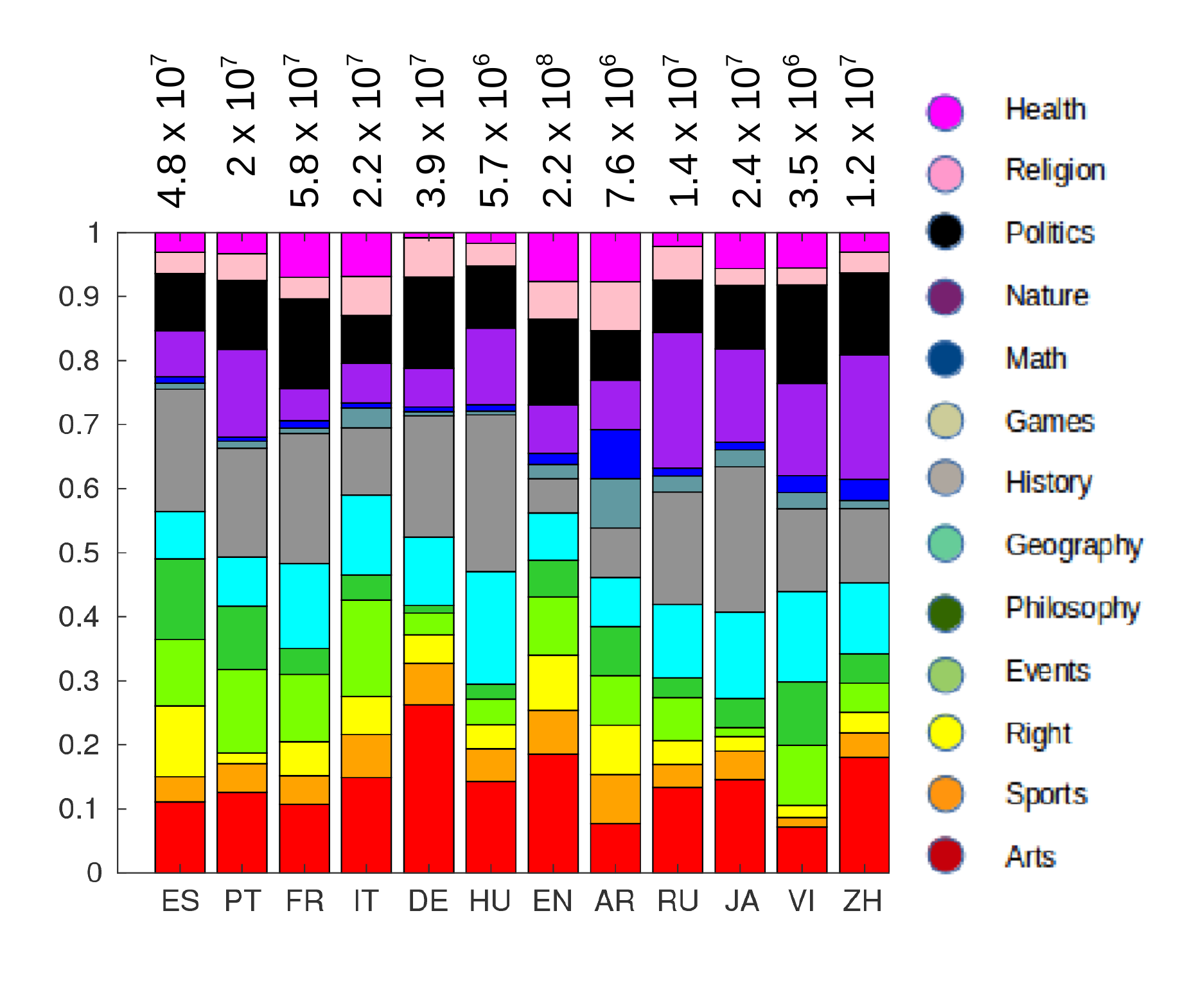}} 
\caption{Distribution of edits for each language. A colour is associated to each category.
The colouring of each column gives the proportion of each category with respect to the total number of edits for the given language.}
\label{fig1}
\end{figure}

We start by showing the number of edits in Fig.~\ref{fig1}. In the edits, we include any action done by WP-users, including correction, addition, and deletion. The motivation for 
including all these actions is the hypothesis stating that the higher the number of edits, the greater the interest on that subject \cite{gandica}. The Y-axis represents, in different 
colors, the proportion of edits in each category. The x-axis shows the language
at the bottom and the total number of edits at the top. Naturally, the EN-WP surpasses the other languages. FR-WP and ES-WP follow as the next more edited languages. The most 
edited categories are `Arts' and `History', followed by `Nature' and `Politics'. All the languages follow roughly similar patterns, with some interesting particularities. For 
example, the category `Arts' is predominantly edited in the DE-WP. While `History' dominates HU-WP and JA-WP. `Nature' appears more important for RU-WP, and `Politics' is the most 
edited category in VI-WP. 

\begin{figure}[tb]
\hspace{-0.5cm}\scalebox{0.48}{\includegraphics{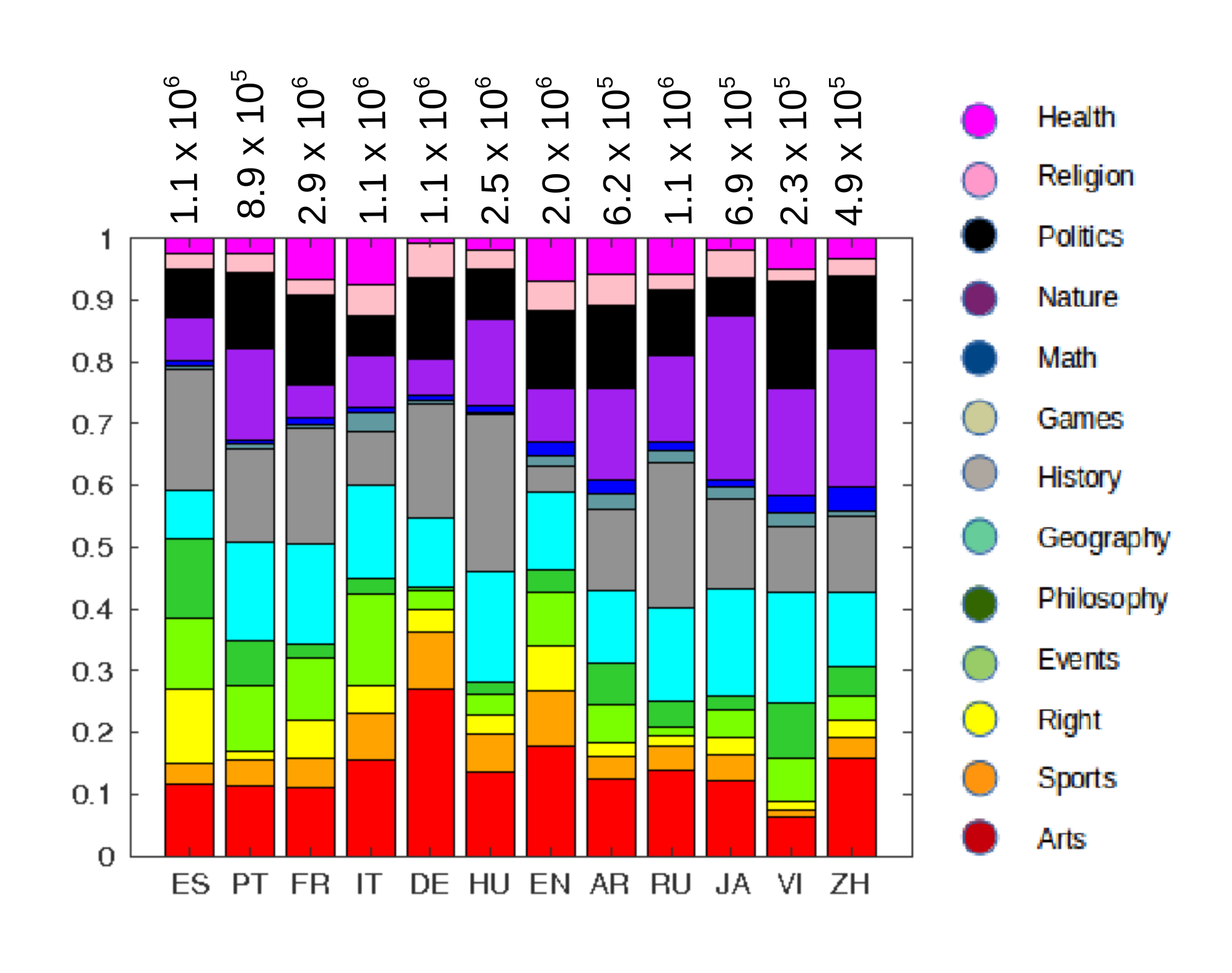}} 
\caption{Distribution of pages for each language. A colour is associated to each category.
The colouring of each column gives the proportion of each category with respect to the total number of pages for the given language.}
\label{fig2}
\end{figure}

It is important to notice that the number of edits, apart from being the result of the collaborative process of editing in order to complement the information, can be highly influenced
by the disagreement between editors, which can lead to edit wars, characterized by a large number of edits \cite{yg}. In this sense, another interesting observable is the total number of 
pages for each language, which is shown in Fig.~\ref{fig2}. Unexpectedly, the highest number of pages is not
in the EN-WP but in the FR-WP. This is an example of the fact that even though the EN-WP is the largest edition, several articles about local places and events, are mostly written only 
in the local languages of those locations \cite{understanding}.
\\

The categories with more WP pages are `Arts', `History' and `Nature'. `Arts' continues to be dominated by DE-WP. But now `History' seems predominant for HU-WP and RU-WP. Then, we can 
hypothesize that the predominant editing behavior in JA-WP could be highly influenced by edit wars. This hypothesis is an interesting point
to be studied in a future work. JA-WP seems to be more interested in developing subjects regarding `Nature' (also taking into account the results for the number of editors). VI-WP 
continues focused on `Politics', this last pattern seeming robust.

\begin{figure}[tbp]
\hspace{-0.5cm}\scalebox{0.48}{\includegraphics{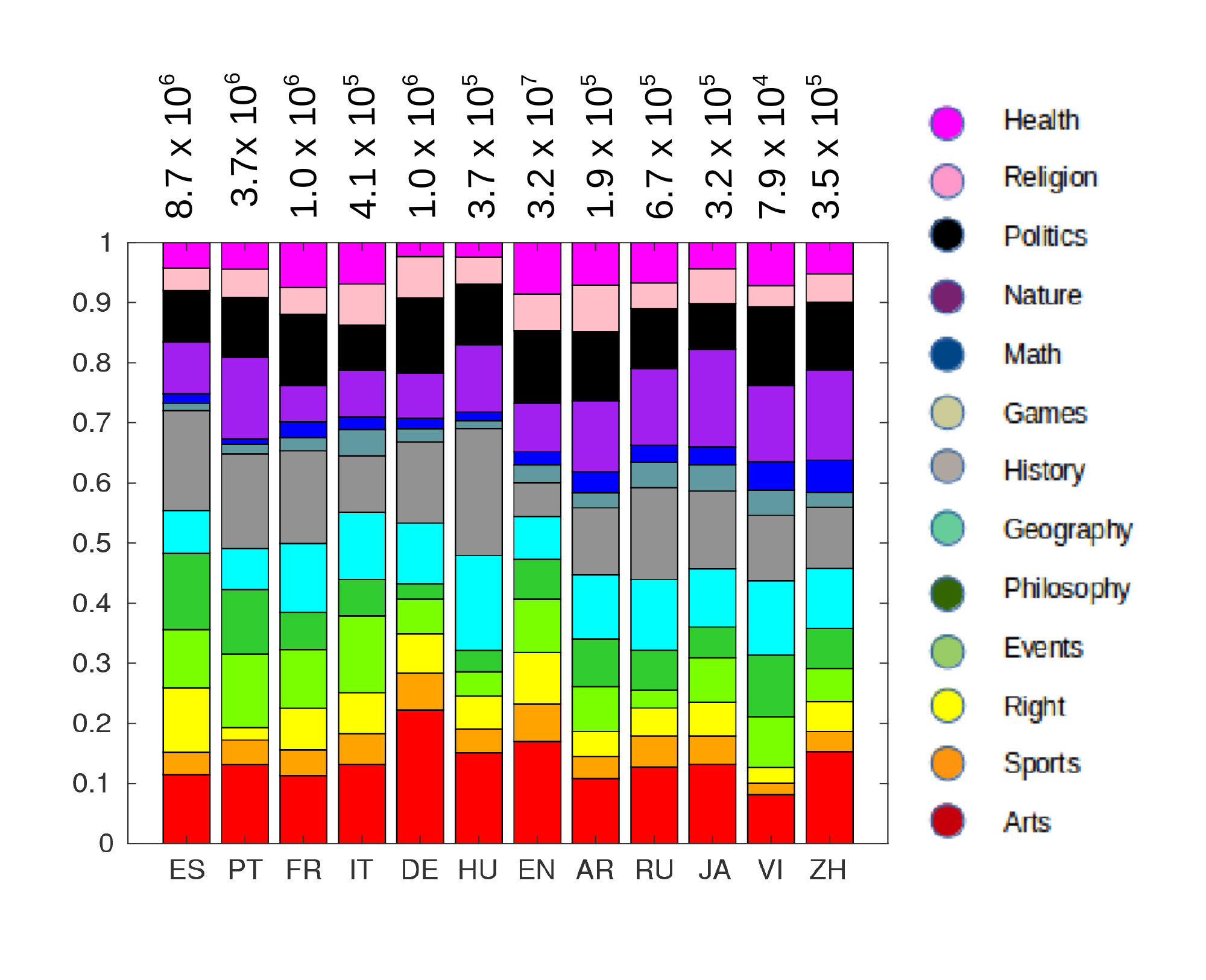}} 
\caption{Distribution of editors for each language. A colour is associated to each category.
The colouring of each column gives the proportion of each category with respect to the total number of editors for the given language.}
\label{fig3}
\end{figure}

We now show in Fig.~\ref{fig3} the number of editors for each language. As in the number of edits, the biggest categories are `Arts' and `History', followed by `Nature' and `Politics'.
The effect of colonialism starts to be visible. ES-WP and PT-WP appear as the most populated. This last calculation confirms the dominant interest on art by DE-WP, while the population 
editing in HU-WP is more interested in `History'.   

\subsection{Homogeneity of the preferences}
\begin{figure}[htbp]
\scalebox{0.35}{\includegraphics{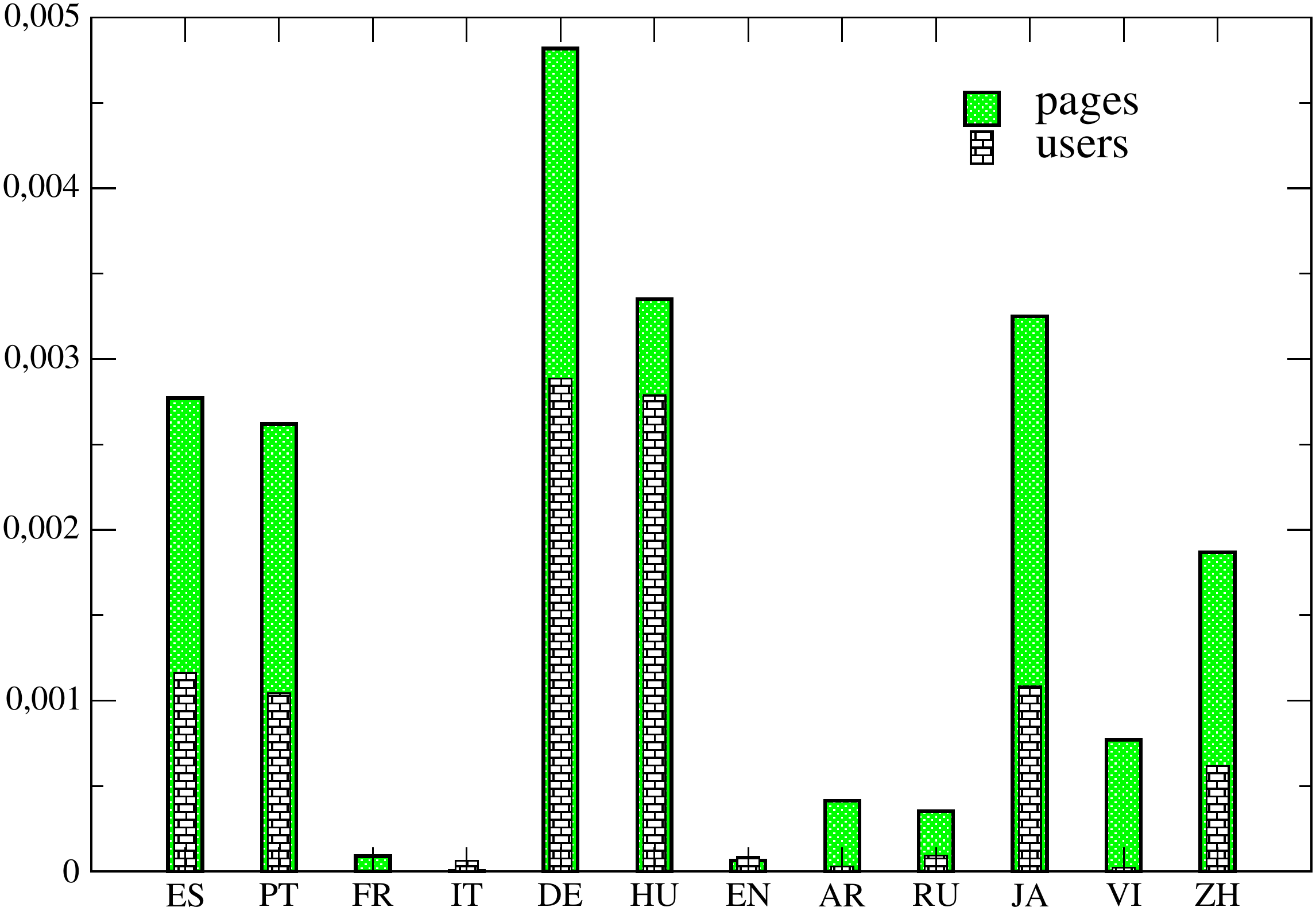}} 
\caption{Degree of homogeneity on the number of pages (in green) and the number of users (in black) among the categories. Calculated by the $\chi^2$-test.}
\label{fig4}
\end{figure}

The degree of homogeneity among the categories, in terms of the number of pages and editors was calculated by means of the $\chi^2$-test with respect to the
uniform distribution. In this sense, the lower the value, the more homogeneous the category distribution will be for that language. Results are reported in Fig.~\ref{fig4}. Our limit case,
the EN-WP, has low values for both, for the number of pages and for the number of editors, probably due to the expected diversity associated with worldwide editing. Surprisingly, those low values are
comparable with the ones by IT-WP and FR-WP. Since the first hypothesis for the homogeneity of preferences is a heterogeneous distribution of people from several nationalities 
--- background---, it is not easy to give an explanation for the last result. In the other extreme, we have DE-WP and HU-WP with the highest variability, as can be noticed in figures 
\ref{fig2} and \ref{fig3}, following local patterns of pronounced preferences.  

\subsection{By categories}

\begin{figure*}[htbp]
\scalebox{0.6}{\includegraphics{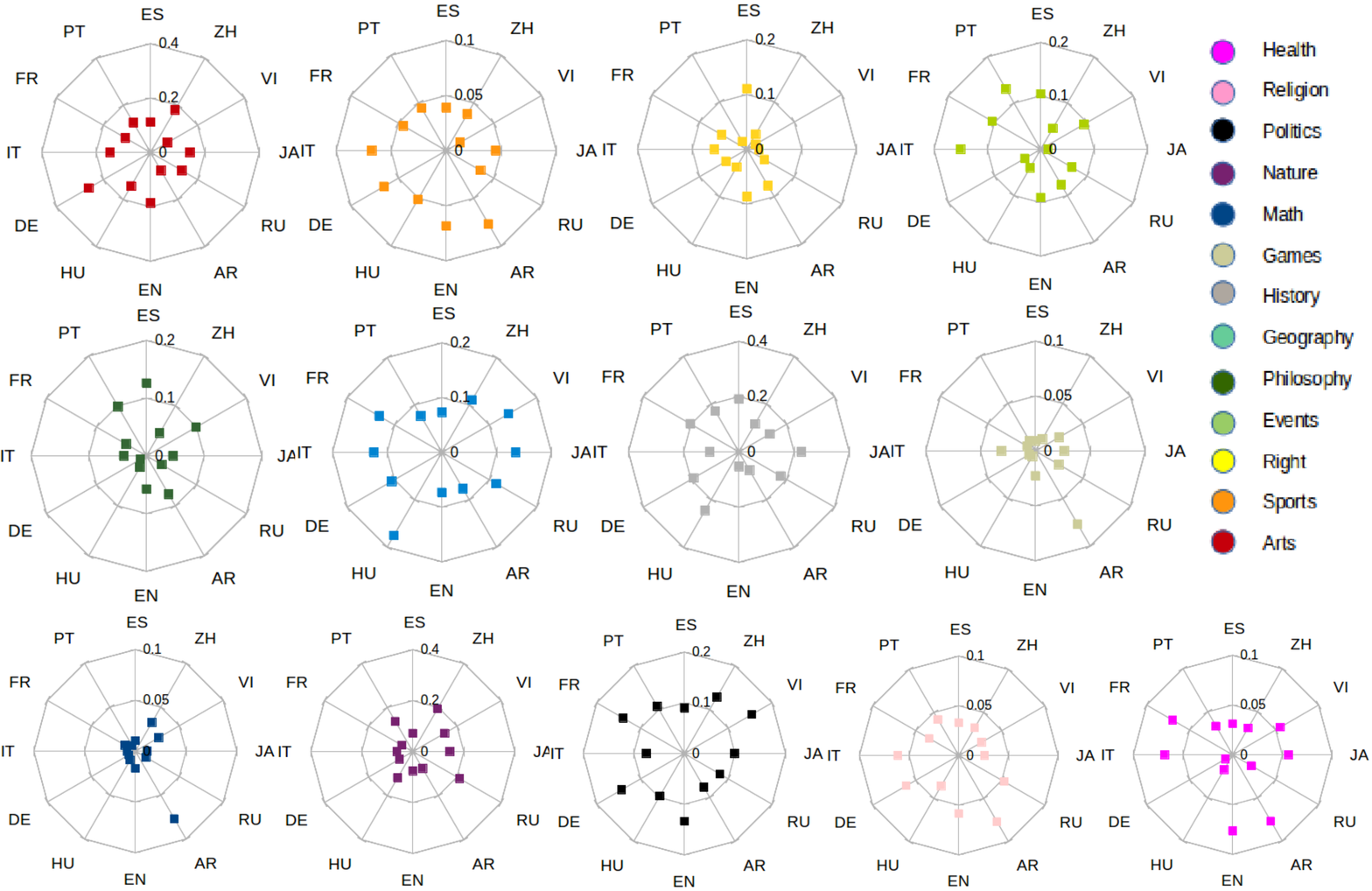}}                
\caption{Each circle represents the proportion of edits among all the WP languages in each category. The colors in each circle represent a category, as indicated in the legend.}
\label{fig5}
\end{figure*}

\begin{figure*}[htbp]
\scalebox{0.6}{\includegraphics{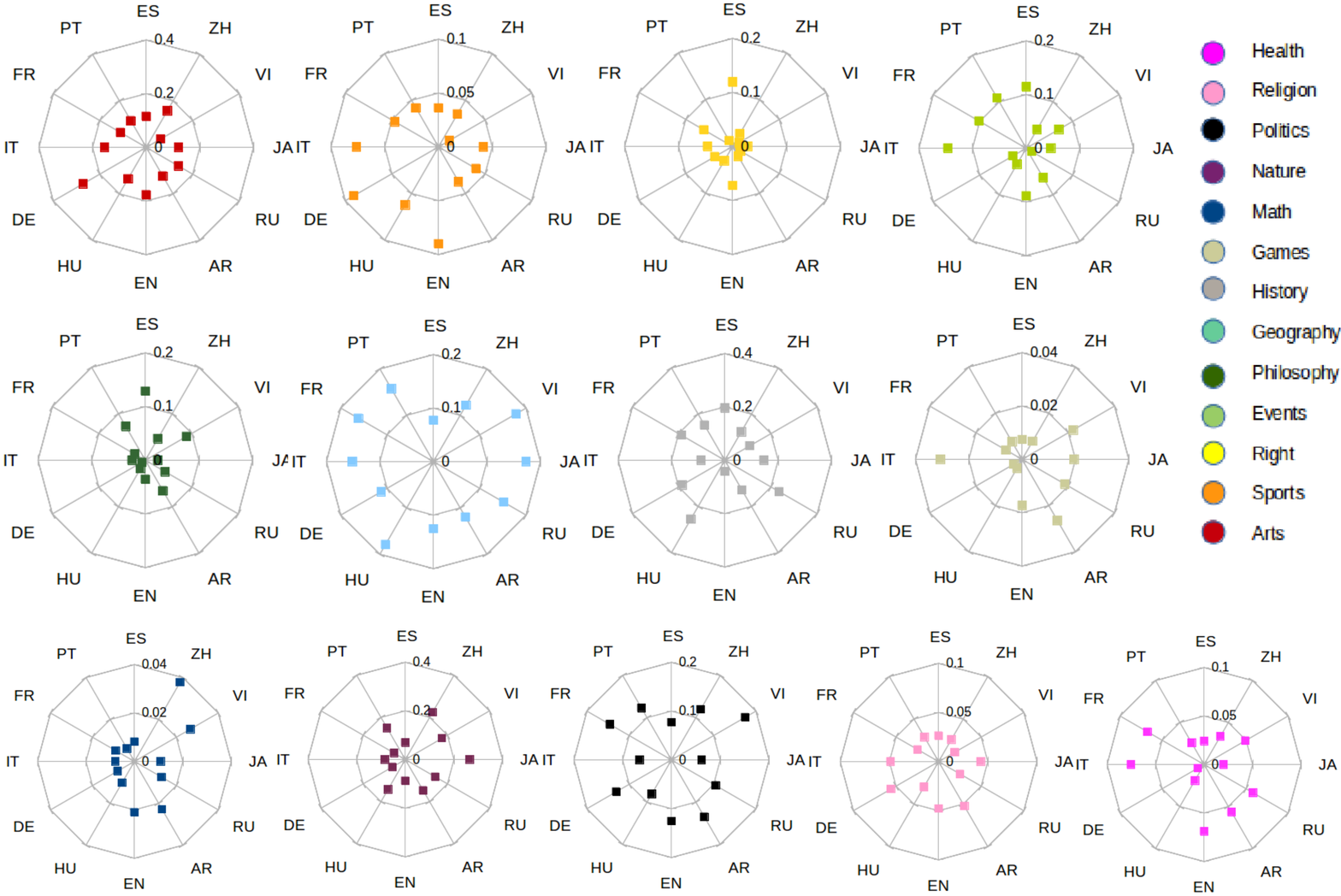}} 
\caption{Each circle represents the proportion of pages among all the WP languages in each category. The colors in each circle represent a category, as indicated in the legend.}
\label{fig6}
\end{figure*}

\begin{figure*}[htbp]
\scalebox{0.6}{\includegraphics{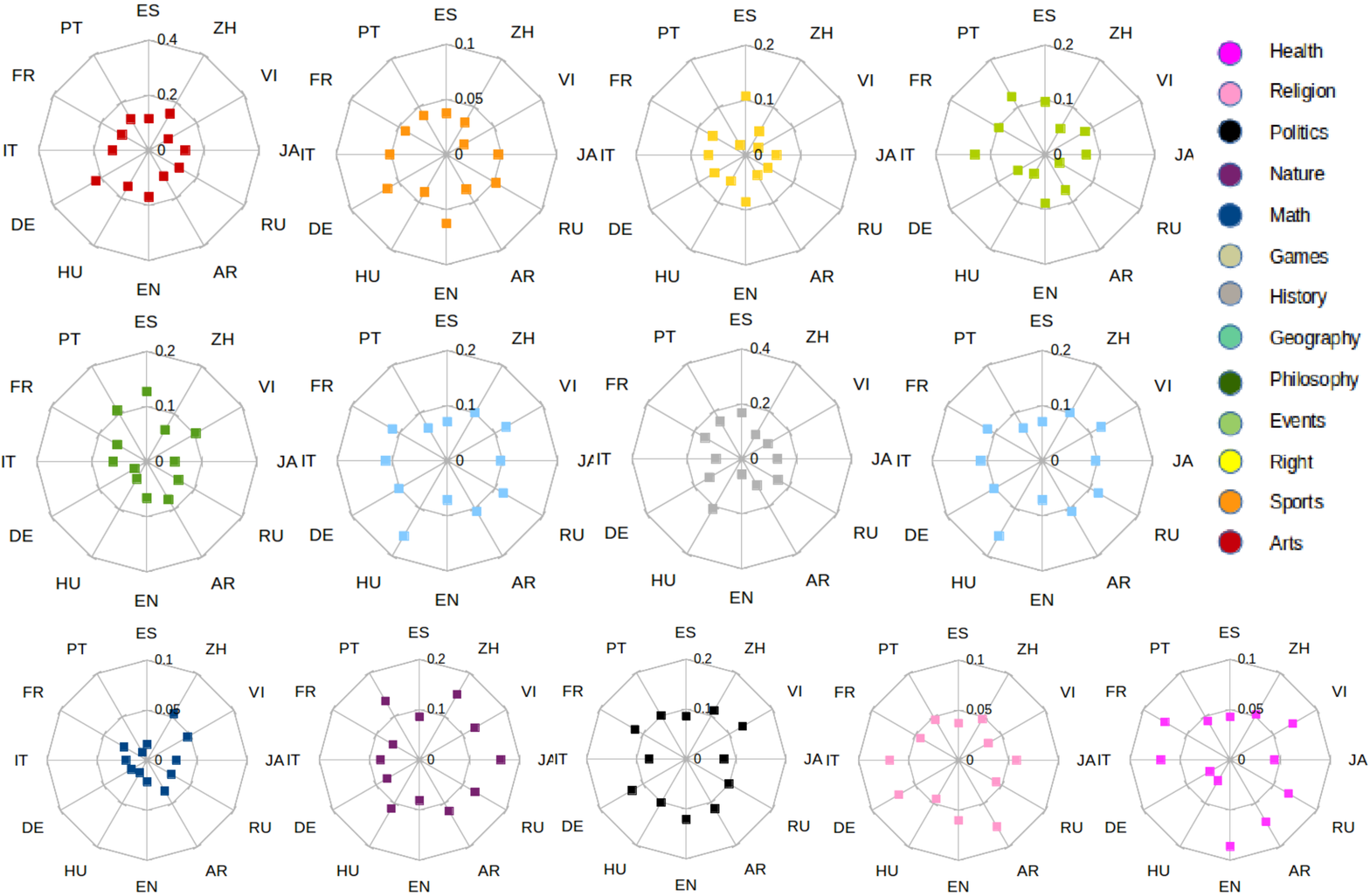}} 
\caption{Each circle represents the proportion of editors among all the WP languages in each category. The colors in each circle represent a category, as indicated in the legend.}
\label{fig7}
\end{figure*}

Now let us explore the results in terms of the categories. How is the proportion for the number of edits, pages and editors in each WP language? The aggregation of all the values
will not be equal to one. However, looking at each value separately gives information about the real proportion in that language and can be compared with the proportion in the other
ones. As we will see the values are clearer in this representation. Results are depicted in Fig.~\ref{fig5},\ref{fig6} and \ref{fig7}, for the number of edits, pages and editors, respectively. Looking at the first icon in the three figures makes it clear 
that DE-WP dominates in the category `Arts', followed by ZH-WP. The representation by languages helps to uncover second and third places, which are relevant as population's 
preferences. The category `Sports' is mostly dominated by DE-WP and IT-WP. `Right' and `Philosophy' are singularly dominant for ES-WP.
The category `Geography' and `History' seem signally important for HU-WP. Taking into account the number of pages and editors, the role of ZH-WP and WI-WP looks important for the 
development of category `Mathematics' and JA-WP for `Nature'. For `Politics' it is remarkable the low activity by IT-WP and JA-WP, while a high activity by VI-WP. Finally, the category 
`health' seems to be more important for IT-WP and FR-WP, in comparison with the rest. 

\section{Limitations}
The main limitation of working with Wikipedia data is the selection-bias, as all the works assume that the proportion of people editing WP is representative of the whole
group under study. The second limitation is the language-based comparisons for cultures, being language only one of the dimensions of the rich and complex elements composing culture
in society. Both limitations are present in all current-state-of-the-art studies. However, it is still possible to gather relevant information from
this data source despite such limitations \cite{mining,cul}. 
\\
\\
Further limitations, specific to our work, include the fact that the categorization could be differently defined among the WP languages. Another limitation is that the same WP page
could belong to several categories. We have tried to reduce the last limitation by leaving out from the study the categories: Culture, Humanities, Law, Life, Matter,
People, Reference Works, Science and Technology, Society, Universe, and World. Such categories should be included whenever the overlap is not an undesirable situation in the
study.
\section{Discussion}

Collective interests of a language-speaking community have been studied by means of the voluntary process of editing. The $12$ languages that have been studied here have 
been selected based on the interplay between a worldwide view and their Wikipedia editing sizes. The categories were taken from the main 
branches of the Wikipedia's tree structure, defined by the same Wikimedia Fundation. We have shown the number of edits, pages and editors for each language, separately
by categories.
\\
\\
We have found stable patterns of preferences by languages, through the number of edits, pages and editors. Some results were unexpected. We were not able to find differences regarding languages spoken in 
one country with respect to the ones spoken in several countries. A contrasting example was the case of the Italian language, which shows characteristics of diversity comparable to the
EN-WP. However, that could be due to the high number of Italian citizens living around the word.
\\
\\
To know the human preferences by languages could help to match data from different languages. The weight of preferences by categories could give some insights about how to perform a 
better match when doing three kinds of analysis among different languages: Machine Translation, Speech Recognition and Sentiment Analysis; All those fields belong to Natural Language Processing. The present work 
is a starting point for that purpose. We hope that our contribution can inspire future attempts at using the differences of human preferences by categories when performing 
Multilingual NLP.
\\
\\
Interesting extensions of our work include discarding the effect of edit wars between wikipedians, in order to have more precise results. Also, the possibility of obtaining
the results for each country, instead of for each language, would be a great contribution. The study of the cultural importance of languages along the lines of the
study of the global influence that each language produces over the rest, presented in \cite{hidalgo_pnas}, would be an important development.
\\
\section{Acknowledgment}
Computational resources have been provided by the Consortium des équipements de Calcul Intensif (CBI), funded by the Fonds de la Recherche Scientifique de Belgique 
(F.R.S.-FNRS) under Grant No.2.5020.11. YG thanks to J\'{e}r\^{o}me Kunegis and Silvia Chiachiera for valuable discussions and technical advises. YG thanks Julieta Barba by technical 
and grafical support. YG thanks to http://www.opensym.org/os2018/ for a previous round of reviewing. YG thanks Fernando Sampaio Dos Aidos for proofreading the article.

\end{document}